\documentclass[sigconf]{acmart}

\usepackage{amsmath}
\usepackage{algorithm}     
\usepackage{algpseudocode}
\usepackage{enumitem}
\usepackage{float}
\usepackage{booktabs} 
\usepackage{multirow}

\usepackage{booktabs} 
\usepackage{colortbl} 
\usepackage{xcolor}   
\usepackage{threeparttable}
\usepackage{caption}

\usepackage{adjustbox}
\usepackage{graphicx}
\usepackage[table,xcdraw]{xcolor}
\usepackage{amsmath}
\usepackage{caption}
\usepackage{makecell}
\usepackage{enumitem}
\usepackage{booktabs}

\usepackage{tabularx}
\usepackage{array}

\AtBeginDocument{%
  }

\setcopyright{acmlicensed}
\copyrightyear{2018}
\acmYear{2018}
\acmDOI{XXXXXXX.XXXXXXX}
\acmConference[Conference acronym 'XX]{Make sure to enter the correct
  conference title from your rights confirmation email}{June 03--05,
  2018}{Woodstock, NY}
\acmISBN{978-1-4503-XXXX-X/2018/06}

\begin{document}

\title{Joint Behavior-guided and Modality-coherence Conditional Graph Diffusion Denoising for Multi Modal Recommendation}

  

\author{Xiangchen Pan}
\affiliation{%
  \institution{Huazhong University of Science and Technology}
  \city{Wuhan}
  \country{China}}
\email{pxcstart666@gmail.com}

\author{WeiWei}
\authornote{Corresponding author.}
\affiliation{%
  \institution{Huazhong University of Science and Technology}
  \city{Wuhan}
  \country{China}
}
\email{weiw@hust.edu.cn}

\renewcommand{\shortauthors}{Trovato et al.}


\begin{abstract}
  In recent years, multimodal recommendation has received significant attention and achieved remarkable success in GCN-based recommendation methods. However, there are two key challenges here: (1) There is a significant amount of redundant information in multimodal features that is unrelated to user preferences. Directly injecting multimodal features into the interaction graph can affect the collaborative feature learning between users and items. (2) There are false negative and false positive behaviors caused by system errors such as accidental clicks and non-exposure. This feedback bias can affect the ranking accuracy of training sample pairs, thereby reducing the recommendation accuracy of the model.

  To address these challenges, this work proposes a Joint Behavior-guided and Modal-consistent Conditional Graph Diffusion Model (JBM-Diff) for joint denoising of multimodal features and user feedback. We design a diffusion model conditioned on collaborative features for each modal feature to remove preference-irrelevant information, and enhance the alignment between collaborative features and modal semantic information through multi-view message propagation and feature fusion. Finally, we detect the partial order consistency of sample pairs from a behavioral perspective based on learned modal preferences, set the credibility for sample pairs, and achieve data augmentation. Extensive experiments on three public datasets demonstrate the effectiveness of this work. Codes are available at https://github.com/pxcstart/JBMDiff. 
\end{abstract}


\maketitle

\section{Introduction}

Recommender systems (RS) are crucial for addressing information overload~\cite{luo2022hysage,luo2022personalized, luo2022towards, luo2024perfedrec++}. In recent years, multimodal recommendation has received increasing attention~\cite{wu2022survey, zhou2023comprehensive}. By incorporating multimodal content information, it is possible to alleviate data sparsity in behavioral interactions, better mine user preferences, and thereby enhance the recommendation performance of the system.

Early multimodal recommendation primarily employed feature-based methods, expanding matrix factorization by incorporating visual features, such as Visual Bayesian Personalized Ranking (VBPR)~\cite{he2016vbpr} and Cross-Modal Knowledge Embedding (CKE)~\cite{zhang2016collaborative}. Later, graph neural networks (GNNs), which can effectively utilize high-order information, were widely applied to mine potential collaborative information between users and items, enhancing user preference representations. Many scholars began to consider enhancing multimodal recommendation using GNNs~\cite{liu2023multimodal,wei2020graph,wei2019mmgcn,xu2018graphcar}. Such work mainly involves constructing a modality-aware auxiliary graph structure to transform multimodal knowledge into item and user embeddings or integrating multimodal features into the user-item interaction graph, aiming to prune false positive interactions using modality information and  add false negative interactions to optimize the interaction graph structure. Although these methods have significantly improved recommendation performance, there are still two key challenges to be addressed.

\textbf{Challenge 1:} Multimodal information contains a significant amount of noise unrelated to user preferences, such as redundant text descriptions and image backgrounds. Directly injecting multimodal features into the interaction graph can affect the collaborative embed learning between users and items. Furthermore, as message propagation deepens, the impact will continue to expand. 

\textbf{Challenge 2:} There are false negative and false positive behaviors caused by system errors such as accidental clicks and non-exposure. Although the introduction of multimodal information can serve as a supervisory screening function on the graph structure, the accuracy of the ranking relationship between positive and negative sample pairs in the training set is the key to influencing model performance. Previous graph recommendation denoising efforts have overlooked the impact of feedback noise on the partial order relationship.

To address the aforementioned challenges, this work proposes a conditional graph diffusion model, JBM-Diff, based on behavior guidance and modality consistency, for joint denoising of multimodal features and behavior feedbacks. The model primarily consists of three components: behavior-guided multimodal denoising, multi-view message propagation and feature fusion, and modality-coherence based behavior debiasing. Firstly, we design a corresponding diffusion model for each modal feature, using the corresponding collaborative feature as the condition for the diffusion model, aiming to remove redundant information unrelated to user preferences. Subsequently, under each modality, we employ the k-nearest neighbor idea to establish a corresponding item-item affinity graph, obtain global modal representations of users and items through a message propagation mechanism, and fuse the modal features to enhance the representation of collaborative features. Finally, we utilize multimodal features to mine user preferences across different modalities, detect the partial order accuracy of positive and negative sample pairs in behavioral view, and perform data augmentation by setting confidence scores, aiming to mitigate feedback noise caused by systematic errors.

In summary, the contributions of this work are primarily manifested in the following three aspects: 
\begin{itemize}[leftmargin=*] 
    \item We propose a conditional graph diffusion model based on behavior guidance and modal consistency, aiming to jointly denoise multimodal features and behavior feedback. 
    \item To remove redundant information unrelated to user preferences from multimodal features, we introduce a behavior-guided conditional diffusion model for each modality; to mitigate the impact of false negative or false positive behaviors on recommendation accuracy, we utilize multimodal features to detect the partial order accuracy and set credibility for sample pairs to achieve data augmentation.
    \item We conducted extensive experiments on three open-source datasets, demonstrating the effectiveness of our work.
\end{itemize}


\section{Related Work}

\subsection{Multi-Model Recommendation}
Multimodal information is ubiquitous in the interaction between recommendation systems and users, playing a crucial role in user decision-making. In recent years, various works have explored incorporating multimodal information into user preference modeling. Early works such as VBPR~\cite{he2016vbpr} extended matrix factorization by integrating product embeddings and multimodal features. In recent years, with the widespread application of graph neural networks in recommendation systems~\cite{he2020lightgcn, fan2019graph, zhao2022multi}, many works have extracted multimodal features through graph neural networks and then fused them with behavioral features to better model user preferences. For example, MMGCN~\cite{wei2019mmgcn}, GRCN~\cite{wei2020graph}, and SLMRec~\cite{tao2022self}. The main modeling process involves first constructing an item-item affinity graph, and then aligning and fusing collaborative information from the user-item interaction graph and semantic information from the item-item affinity graph through different methods. For instance, MICRO~\cite{zhang2022latent} designed a contrastive task to promote multimodal fusion and optimize item representations; FREEDOM~\cite{zhou2023tale} froze the item-item multimodal graph and denoised the user-item graph through edge pruning. However, many multimodal recommendation works have overlooked the inherent noise in these modal features. This work focuses on how to remove redundant information unrelated to user preferences from multimodal features.

\subsection{Diffusion Models in Recommendation}
Inspired by the successful applications of diffusion models (DMs) in image synthesis~\cite{rombach2022high}, text generation~\cite{li2022diffusion}, and other fields, some studies have explored the effectiveness of DMs in recommendation. For example, DiffRec~\cite{wang2023diffusion} gradually generates globally similar yet personalized collaborative signals through the denoising process of DMs; DreamRec~\cite{yang2023generate} utilizes DMs to explore the underlying distribution of the recommended item space and generate guidance for user sequential behavior. In addition, in recent years, DMs have also been widely applied in multimodal recommendation. For instance, DiffMM~\cite{jiang2024diffmm} integrates a modality-aware graph model and cross-modality contrastive learning and enhances the structure of the item-item affinity graph in each modality through DMs; MCDRec~\cite{ma2024multimodal} takes multimodal features as conditions and enhances collaborative information on the item side through a conditional diffusion model, which is then used for denoising and pruning of the user-item graph. This work attempts to use DMs and leverage behavioral features as conditions to guide feature enhancement for multimodal information and remove modality noise.

\subsection{Graph-based Recommendation}
Graph neural networks have been widely applied to recommendation systems to capture potential high-order relationships from user interaction behavior data. GCN-based recommendation methods treat interaction data as bipartite graphs and use graph neural networks to model high-order transformations between users and items, such as NGCF~\cite{wang2019neural}, GCCF~\cite{sun2019multi}, LightGCN~\cite{he2020lightgcn}, etc. However, research~\cite{gao2022self, wang2021denoising, wang2022learning} has shown that user interaction behavior data inevitably contains noise, such as accidental clicks and popularity bias. To address these challenges, researchers have attempted to design contrastive learning tasks through self-supervised learning to enhance the robustness of model recommendations, such as SGL~\cite{wu2021self}, SimGCL~\cite{yu2022graph}, AutoCF~\cite{xia2023automated}, etc. However, recommendation algorithms primarily use binary cross-entropy loss as the main objective for training, and the partial order accuracy of training sample pairs is the key factor affecting recommendation performance. Therefore, this work attempts to utilize multimodal preferences to supervise the partial order consistency of positive and negative samples by setting confidence weights for each sample to mitigate the impact of behavioral noise on recommendation accuracy from the data level.

\section{Methodology}

\begin{figure}[t]
    \centering
    \includegraphics[width=1\linewidth]{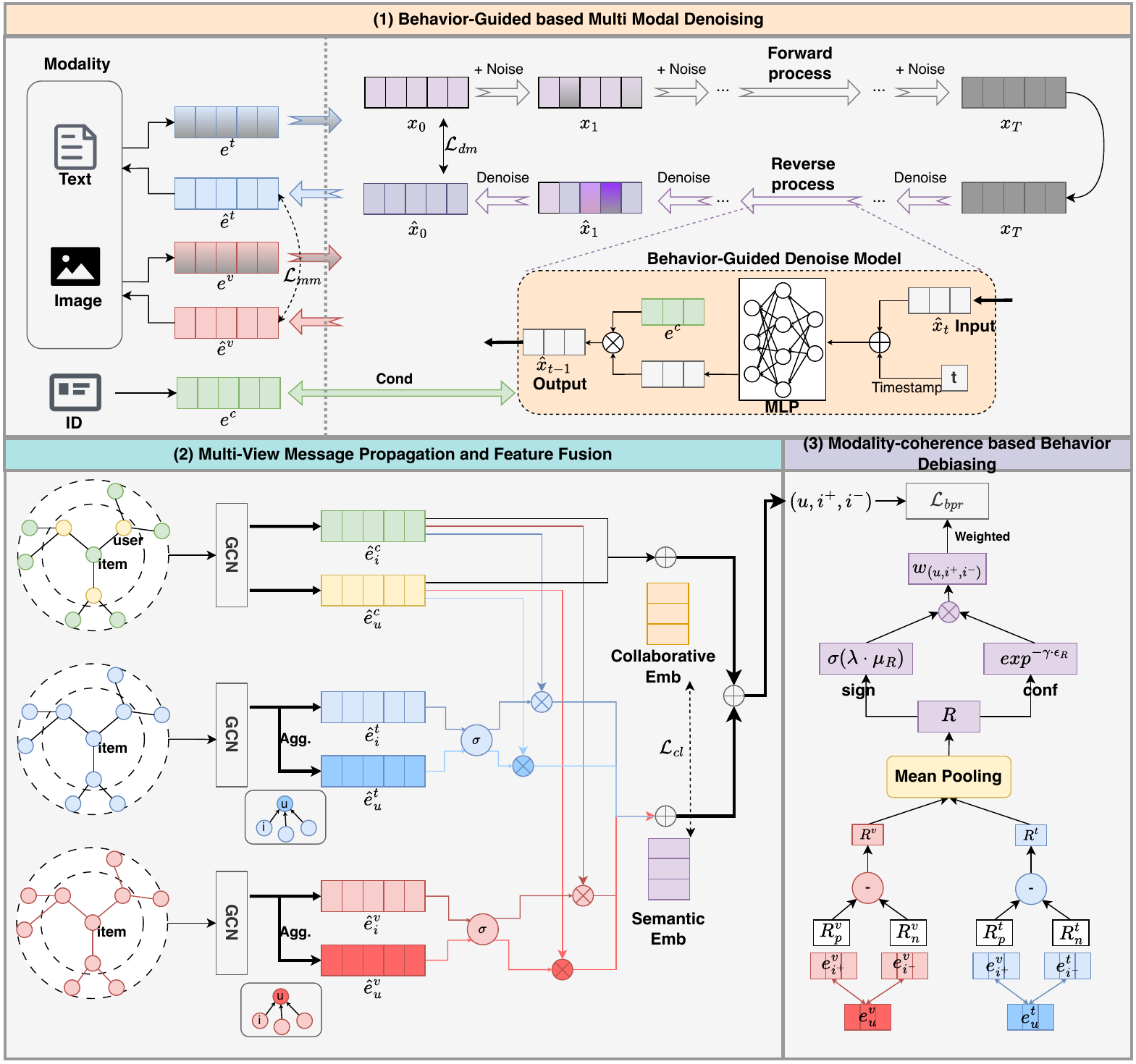}
    \caption{The overview framework of JBM-Diff. It consists of three major components, i.e., behavior-guided Multi Modal Denoising, Multi-View Message Propagation and Feature Fusion, and Modality-coherence based Behavior Debiasing.}
    \label{overview}
\end{figure}

In this section, we will introduce the Joint Behavior-guided and Modal-consistent Conditional Graph Diffusion Model (JBM-Diff) and how it performs joint denoising on multimodal and behavioral features. As shown in Fig. ~\ref{overview}, JBM-Diff mainly consists of three parts: considering the redundant information unrelated to user preferences in multimodal data, we propose a multimodal denoising method based on behavior guidance, designing corresponding conditional denoising models for each modal feature to enhance the representation of modal features (Section 3.2); to further utilize multimodal features to enhance user preference representation, we adopt multi-view semantic modeling, constructing corresponding item-item graphs for each modality based on the k-nearest neighbor principle, to comprehensively capture semantic information under different views (Section 3.3); to mitigate the impact of behavior noise on recommendation accuracy, we propose a behavior debiasing scheme based on modality consistency, which comprehensively evaluates preference rankings under each modality, checks the consistency of preference orders between sample pairs on the collaborative side, sets credibility for each sample pair, and achieves behavior debiasing through data augmentation (Section 3.4).


\subsection{Preliminaries}
Let $U$ and $I$ denote the user and item sets, respectively. $|U|$ and $|I|$ represent the total number of users and items, respectively. The user-item interaction matrix is denoted as $O \in R^{|U|*|I|}$. Here, $O_{u,i}=1$ indicates that the user $u$ has interacted with item $i$. The features of the item $i$ for each modality $m$ have been preprocessed and set to $e_i^m \in R^{d_m}$, where $d_m$ is the dimension of the feature for modality $m$. In this work, the modality type can be $m \in M=\{v,t\}$, where $v$ represents visual information and $t$ represents textual information. The goal of the recommendation task is to predict the preference score $y_{ui}$ for a given user and item based on $O$ and $\{e_m\}_{m \in M}$.


\subsection{Behavior-guided Multi Modal Denoising}
To remove redundant information unrelated to user preferences from multimodal features, we propose a behavior-guided multimodal denoising method. Inspired by the successful application of Diffusion Models (DMs) in computer vision and natural language processing, we attempt to apply the DM model to enhance multimodal features in recommendation scenarios. Unlike other diffusion-based recommendation works, we do not enhance the graph structure but directly enhance the multimodal features on the item side. We also utilize the collaborative features of items as conditions to guide modal denoising. Below, we will specifically introduce the forward and reverse processes of the diffusion model.


\subsubsection{Forward Process}
We establish a corresponding diffusion model for each modality $m$. Without loss of generality, we denote the modality feature $e_i^m$ of the item $i$ as the initial state $x_0^m$. The forward process constitutes a Markov chain, gradually adding Gaussian noise to the original data samples over $T$ time steps $t \in {1,...,T}$. This process ultimately produces a noise vector $x_T^m$ that approximates a standard normal distribution. Each forward transition step is defined as:


\begin{equation}
\begin{aligned}
q(x_t^m | x_{t-1}^m) = \mathcal{N}(x_t^m; \sqrt{1-\beta_t^m}x_{t-1}^m,\beta_t^mI) 
\end{aligned}
\end{equation}

Where $\beta_t^m \in (0,1)$ is the Gaussian noise scale at time step $t$, and $I$ is the identity matrix. According to the reparameterization trick, we can converge the noise process at each step. Let $\alpha_t^m=1-\beta_t^m$, and define the cumulative product as $\bar{\alpha}_t^m=\Pi_{t'=1}^t \alpha_{t'}^m$. Therefore, the noise sample $x_t^m$ at the step $t$ can be directly derived from the original state $x_0^m$ as follows:

\begin{equation}
\begin{aligned}
q(x_t^m | x_0^m) &= \mathcal{N}(x_t^m; \sqrt{\bar{\alpha}_t^m}x_t^m,(1-\bar{\alpha}_t^m)I) \\
x_t^m &= \sqrt{\bar{\alpha}_t^m}x_0^m + \sqrt{1-\bar{\alpha}_t^m}\epsilon
\end{aligned}
\end{equation}

where $\epsilon \in \mathcal{N}(0,I)$.

\subsubsection{Reverse Process}
The reverse process of the diffusion model aims to iteratively denoise $x_t$ through a T-step denoising process, ultimately obtaining an approximate original representation $\hat{x}_0$. Here, we use the collaborative feature $e_i^c$ of the item $i$ as a condition and guide the removal of redundant information unrelated to preferences from multimodal features through the denoising model $f_{\theta}$. The conditional generation process at each step $t$ can be represented as follows:

\begin{equation}
\begin{aligned}
p_\theta(x_{t-1}^m|x_t^m, e^c)=\mathcal{N}(x_{t-1}^m; \mu_\theta(x_t^m,t,e^c),\Sigma_{\theta}(x_t^m,t, e^c))
\end{aligned}
\end{equation}

Where $\mu_\theta$ and $\Sigma_\theta(\cdot)$ represent the learnable mean and covariance predicted by the network. The relevant calculation formula is as follows:

\begin{equation}
\begin{aligned}
\mu(x_t^m, t, e^c) &= \frac{1}{\sqrt{\alpha_t}}(x_t^m - \frac{\beta_t}{\sqrt{1-\bar{\alpha}_t}}f_\theta(x_t^m,t,e^c)) \\
\Sigma_\theta(x_t, t, e^c) &= \frac{1-\bar\alpha_{t-1}}{1-\bar{\alpha}_t}\beta_t I
\end{aligned}
\end{equation}

Regarding the denoise model $f_{\theta}$, we adopt an MLP architecture. First, we concatenate the multimodal features $x_t^m$ and the time-step features $e_t$, then map them to the behavior space, and perform element-wise dot product operations with the collaborative features $e^c$ of the item to obtain the enhanced $x_{t-1}^m$:

\begin{equation}
\begin{aligned}
f_\theta(x_t^m,t,e^c) = MLP_\theta([x_t^m,e_t]) \odot e^c
\end{aligned}
\end{equation}

After $T$ steps of denoising, we can obtain an approximate original representation $\hat{x}_0^m$ as the enhanced result. In order to maintain the personalized characteristics of the item while introducing diffusion multimodal information, we ultimately adopt $\hat{e}_i^m=(1-\omega)x_0^m + \omega \cdot \hat{x}_0^m$ as the feature representation of item $i$ in modality $m$. The mean squared error loss of the diffusion model is expressed as follows:

\begin{equation}
\begin{aligned}
\mathcal{L}_{dm}^m = E_{x_0^m, x_t^m}[||x_0^m-f_\theta(x_t^m,t,e^c)||^2]
\end{aligned}
\end{equation}

In addition, since we have set up corresponding diffusion models for each modality, to ensure consistency in understanding user preferences across modalities, we have also established a self-supervised auxiliary task for inter-modality alignment:

\begin{equation}
\begin{aligned}
\mathcal{L}_{mm} = \frac{1}{|I|}\sum_{i \in I}-log\frac{exp(\hat{e}_i^t \cdot \hat{e}_i^v /\tau)}{\sum_{i' \in I}exp(\hat{e}_i^t \cdot \hat{e}_{i'}^v /\tau)} 
\end{aligned}
\end{equation}

\subsection{Multi-View Message Propagation and Feature Fusion}
Inspired by LightGCN's modeling on high-order collaborative signals for mining user preferences, we adopt a k-nearest neighbor strategy across various modalities to construct an item-item affinity graph. This approach enables us to explore the latent semantic information within each modality from different perspectives, thereby enriching and refining the representation of behavioral features. Next, we will introduce the corresponding message propagation processes from the collaborative and semantic perspectives, as well as how to perform feature fusion.


\subsubsection{User-Item Collaborative View}
To capture high-order interaction information in the user-item interaction graph, we designed an interaction-centered message propagation module composed of GCN. The message propagation of the $l+1$ graph convolutional layer can be formally defined as:

\begin{equation}
\begin{aligned}
E_c^{l+1} = (D^{-\frac{1}{2}}AD^{-\frac{1}{2}})E_c^l
\end{aligned}
\end{equation}

Where $A \in R^{(|U|+|I|)\times (|U|+|I|)}$ is the adjacency matrix constructed based on the interaction matrix $O$, and $D$ is the diagonal matrix of $A$. Each diagonal element $D_{j,j}$ represents the number of non-zero elements in the $j$-th row of matrix $A$. The embedding matrix representation $E_c^0$ is initialized as $E^id$ using the id representations of users and items.


The representation of layer $l$ encodes the information of l-level neighbors. By aggregating the information of higher-order neighbors, the final collaborative representation is obtained:

\begin{equation}
\begin{aligned}
\hat{E}_c = \frac{1}{L+1}\sum_{l=0}^{L}E_c^{(l)}
\end{aligned}
\end{equation}

\subsubsection{Item-Item Semantic View}
Similar to the user-item interaction graph on the collaborative side, we construct an item-item semantic graph for each modality $m$ based on the corresponding item features to capture semantic correlation signals. Considering the large computational cost of propagating modality features on dense graphs and the potential introduction of noise, we use k-nearest neighbors (kNN) for sparsification. Specifically, we calculate the semantic similarity between items based on modality features, and for each node in the graph, we select the top k nodes that are most semantically related to it for connection. The semantic similarity calculation employs cosine similarity:


\begin{equation}
\begin{aligned}
s_{a,b}^m=\frac{(e_a^m)^Te_b^m}{||e_a^m||\cdot||e_b^m||}
\end{aligned}
\end{equation}

The item-item affinity graph's adjacency matrix under modality $m$ is represented as follows:

\begin{equation}
\begin{aligned}
\bar{s}_{a,b}^m=\left\{
\begin{aligned}
& s_{a,b}^m,  s_{a,b}^m \in top-K(s_{a,c}^m, c \in I) \\
& 0, otherwise
\end{aligned}
\right.
\end{aligned}
\end{equation}

Similar to the feature aggregation method on the collaborative side, we use GCN to propagate messages on the item-item affinity graph. However, considering that semantic similarity decreases continuously as the propagation path increases, we only use one layer of graph convolutional network to obtain the modal aggregation representation on the item side:

\begin{equation}
\begin{aligned}
\hat{E}_i^m=(D_m^{-\frac{1}{2}}S^mD_m^{-\frac{1}{2}})E_i^m
\end{aligned}
\end{equation}

Finally, we aggregate the semantic features of the interacted items for each user, obtaining the user representation in that modality:

\begin{equation}
\begin{aligned}
\hat{E}_u^m=\sum_{i \in \mathcal{N}_u}\frac{1}{\sqrt{|\mathcal{N}_u|\cdot|\mathcal{N}_i|}}\hat{E}_i^m
\end{aligned}
\end{equation}

\subsubsection{Feature Fusion}
After aggregating collaborative and semantic features, we introduce a behavior-based gating mechanism to fuse different modal features based on their weights, resulting in the final semantic representation $E_s$, given the varying importance of features across different modalities:


\begin{equation}
\begin{aligned}
E_s = \frac{1}{|M|}\sum_{m \in M}\hat E_m\odot \sigma(W\cdot \hat{E}_c+d)
\end{aligned}
\end{equation}

By adding the semantic feature $E_s$ and the collaborative feature $\hat{E}_c$, we can obtain the final representation of the user and the item, denoted as $E=E_s+\hat{E}_c$. To align the collaborative signal and the semantic signal, we set up a self-supervised contrastive learning task:

\begin{equation}
\begin{aligned}
\mathcal{L}_{cl}=\frac{1}{|U|}\sum_{u \in U}-log\frac{exp(e_u^s\cdot \hat{e}_u^c)}{\sum_{v\in U}exp(e_u^s\cdot \hat{e}_v^c)} + \frac{1}{|I|}\sum_{i \in I}-log\frac{exp(e_i^s\cdot \hat{e}_i^c)}{\sum_{j\in I}exp(e_i^s\cdot \hat{e}_j^c)}
\end{aligned}
\end{equation}

\subsection{Modality-coherence based Behavior Debiasing}
After obtaining the feature representations of users and items, recommendation systems generally adopt personalized Bayesian Probability Ranking Loss (BPR loss) as the primary task for model training. The positive samples in the training sample pairs are the items that the user has interacted with, while the negative samples are often obtained through random sampling. Considering the behavioral noise in the dataset caused by system errors such as false clicks and non-exposures, some sample pairs that appear reasonable from the perspective of behavioral interaction may not truly reflect user preferences. To eliminate the impact of behavioral noise on recommendation accuracy, we calculate the confidence based on the partial order relationship of modal preferences, and then assign weights to each sample.


Specifically, for a sample pair $(u, i_p, i_n)$ where $i_p$ represents a positive sample that has interacted with user $u$, and $i_n$ represents a negative sample that has not interacted with user $u$, we will comprehensively evaluate both \textbf{preference consistency} and \textbf{modal conflict} to score the partial order credibility of this sample pair.


Firstly, under each modality m, the dot product of $i_p$ and $i_n$ with user u is calculated separately to obtain the predicted preference scores $\hat y_{u,i^+}^m$ and $\hat y_{u,i^-}^m$. These scores are then projected onto the probability space using an activation function to represent the user u's preference for the item in modality m. Next, the preference gap between positive and negative samples is calculated for each modality, and the probability differences for each modality are concatenated to obtain the modality bias $R$. This process can be formally defined as follows:


\begin{equation}
\begin{aligned}
R_p^m=\sigma (e_u^m e_{i^+}^m)&; R_n^m=\sigma (e_u^m e_{i^-}^m) \\
R_p = [R_p^{m_1};...;R_p^{m_{|M|}}]&; R_n = [R_n^{m_1};...;R_n^{m_{|M|}}] \\
R=&R_p-R_n
\end{aligned}
\end{equation}

We calculate the mean $sign$ and variance $conf$ for the modal bias $R$. The mean $sign$ reflects the overall degree of ranking bias of the sample pair in modal semantics. The larger the $sign$, the stronger the consistency between behavior and semantic partial order, and correspondingly, the higher the confidence of the sample. The variance $conf$ reflects the degree of difference in preference understanding within different modalities. The larger the $conf$, the more obvious the modal conflict, and correspondingly, the lower the confidence of the sample. The calculation of the confidence $w_{(u,i^+,i^-)}$ for a sample pair is formally defined as follows:


\begin{equation}
\begin{aligned}
w_{(u,i^+,i^-)}&=sign*conf \\
sign = \sigma(\lambda \cdot \mu_R)&; conf=exp^{-\gamma \cdot \epsilon_R}
\end{aligned}
\end{equation}

Where $\mu_R$ and $\epsilon_R$ represent the mean and variance of the modality bias $R$, respectively, $\lambda$ and $\gamma$ are temperature coefficients used to control the relative weights of the two influencing factors: preference consistency and modal conflict. After weighting each sample based on its confidence, the denoising BPR loss is designed as follows:

\begin{equation}
\begin{aligned}
\mathcal{L}_{bpr}=\sum_{(u,i^+,i^-)\in \mathcal{D}}w_{(u,i^+,i^-)}(-\log \sigma(\hat{y}_{ui^+}-\hat{y}_{ui^-})
)\end{aligned}
\end{equation}

\subsection{Model Optimization}
In the model optimization stage, we use denoised BPR loss as the primary learning objective. The overall objective function $\mathcal{L}$ can be expressed as:


\begin{equation}
\begin{aligned}
\mathcal{L}=\mathcal{L}_{bpr} + \lambda_{dm}\cdot\mathcal{L}_{dm} + \lambda_{mm}\cdot\mathcal{L}_{mm} + \lambda_{cl}\cdot\mathcal{L}_{cl}
\end{aligned}
\end{equation}

Among them, $\lambda_{dm}$, $\lambda_{mm}$, and $\lambda_{cl}$ serve as hyperparameters to control the importance of different loss terms.

\section{Experiment}

We evaluated the recommendation performance of our model on real-world datasets and answered the following questions. 
\begin{itemize}[leftmargin=*] 
    \item \textbf{RQ1}: How does our model compare to other baseline models, and whether it is superior to the most advanced models currently available.
    \item \textbf{RQ2}: The impact of each module in our model on overall performance; 
    \item \textbf{RQ3}: The impact of some key hyperparameter settings on model performance; 
    \item \textbf{RQ4}: The recommendation stability of the model in noisy scenarios.
\end{itemize}

\subsection{Experiment Setting}

\subsubsection{Dataset Settings}
Following~\cite{zhang2021mining, zhou2023bootstrap}, we conducted experiments using three publicly available multimodal recommendation datasets, which are sourced from the Baby, Sports, and Clothing subsets of the Amazon Review Dataset\footnote{https://jmcauley.ucsd.edu/data/amazon/links.html}. The statistical information for these datasets can be found in Table ~\ref{table1}.

\begin{table}[ht]
\centering
\caption{Statistics of the experimental datasets}
\label{table1}
\resizebox{0.5\textwidth}{!}{  
\begin{tabular}{lrrrrr}
\toprule
\textbf{Dataset} & \textbf{\# Users} & \textbf{\# Items} & \textbf{\# Interactions} & \textbf{Density}  \\
\midrule
Baby & 19445 & 7050 & 160792 & 0.117\% \\
Sports & 35598 & 18357 & 296337 & 0.045\% \\
Clothing & 39387 & 23033 & 278677 & 0.031\% \\
\bottomrule
\end{tabular}
}
\end{table}

\subsubsection{Baseline}
To demonstrate the effectiveness of our model, we compared JBM-Diff with the following baseline models:

\begin{itemize}[leftmargin=*] 
    \item The first category consists of general collaborative filtering models and traditional graph-based recommendations, such as BPR~\cite{rendle2012bpr}, LightGCN~\cite{he2020lightgcn}, SGL~\cite{wu2021self}, and NCCL~\cite{lin2022improving}; 
    \item The second category involves hypergraph-based recommendations, including HCCF~\cite{xia2022hypergraph}, SHT~\cite{xia2022self}, and LGMRec~\cite{guo2024lgmrec}; 
    \item The third category encompasses contrastive learning-based multimodal recommendations, including VBPR~\cite{he2016vbpr}, MMGCN~\cite{wei2019mmgcn}, GRCN~\cite{wei2020graph}, LATTICE~\cite{zhang2021mining}, MMGCL~\cite{yi2022multi}, MICRO~\cite{zhang2022latent}, SLMRec~\cite{tao2022self}, and BM3~\cite{zhou2023bootstrap}; 
    \item The fourth category features diffusion model-based multimodal recommendations, such as DiffMM~\cite{jiang2024diffmm} and MCDRec~\cite{ma2024multimodal}.
\end{itemize}

\begin{table*}[htbp]
\centering
\caption{Overall performance of JBM-Diff and different baselines in terms of Recall@K(R@K) and NDCG@K(N@K). The best result is bold, and the second best is \underline{underlined}. Improvement percentage (Improv.) denotes the ratio of performance increment from the suboptimal method to JBM-Diff. ‘*’ indicates that the improvement is statistically significant with a p-value < 0.01 by a paired t-test.}
\label{table2}
\resizebox{\textwidth}{!}{
\begin{tabular}{l|cccc|cccc|cccc}
\hline
Datasets 
& \multicolumn{4}{c|}{\textbf{Baby}} 
& \multicolumn{4}{c|}{\textbf{Sports}} 
& \multicolumn{4}{c}{\textbf{Clothing}} \\
\cline{1-13}
Metrics
& R@10 & R@20 & N@10 & N@20 
& R@10 & R@20 & N@10 & N@20 
& R@10 & R@20 & N@10 & N@20 \\
\hline

BPR      & 0.0379 & 0.0607 & 0.0202 & 0.0261 & 0.0452 & 0.0690 & 0.0252 & 0.0314 & 0.0211 & 0.0315 & 0.0118 & 0.0144 \\
LightGCN & 0.0464 & 0.0732 & 0.0251 & 0.0320 & 0.0553 & 0.0829 & 0.0307 & 0.0379 & 0.0331 & 0.0514 & 0.0181 & 0.0227 \\
SGL      & 0.0532 & 0.0820 & 0.0289 & 0.0363 & 0.0620 & 0.0945 & 0.0339 & 0.0423 & 0.0392 & 0.0586 & 0.0216 & 0.0266 \\
NCL      & 0.0538 & 0.0836 & 0.0292 & 0.0369 & 0.0616 & 0.0940 & 0.0339 & 0.0421 & 0.0410 & 0.0607 & 0.0228 & 0.0275 \\
\hline
HCCF     & 0.0480 & 0.0756 & 0.0259 & 0.0332 & 0.0573 & 0.0857 & 0.0317 & 0.0394 & 0.0342 & 0.0533 & 0.0187 & 0.0235 \\
SHT      & 0.0470 & 0.0748 & 0.0256 & 0.0329 & 0.0564 & 0.0838 & 0.0306 & 0.0384 & 0.0345 & 0.0541 & 0.0192 & 0.0243 \\
LGMRec   & \underline{0.0644} & 0.1002 & \underline{0.0349} & \underline{0.0440} & 0.0720 & 0.1068 & 0.0390 & 0.0480 & 0.0555 & 0.0828 & 0.0302 & 0.0371 \\
\hline  
VBPR     & 0.0424 & 0.0663 & 0.0223 & 0.0284 & 0.0556 & 0.0854 & 0.0301 & 0.0378 & 0.0281 & 0.0412 & 0.0158 & 0.0191 \\
MMGCN    & 0.0398 & 0.0649 & 0.0211 & 0.0275 & 0.0382 & 0.0625 & 0.0200 & 0.0263 & 0.0229 & 0.0363 & 0.0118 & 0.0152 \\
GRCN     & 0.0531 & 0.0835 & 0.0291 & 0.0370 & 0.0600 & 0.0921 & 0.0324 & 0.0407 & 0.0431 & 0.0664 & 0.0230 & 0.0289 \\
LATTICE  & 0.0536 & 0.0858 & 0.0287 & 0.0370 & 0.0618 & 0.0950 & 0.0337 & 0.0423 & 0.0459 & 0.0702 & 0.0253 & 0.0306 \\
MMGCL    & 0.0522 & 0.0778 & 0.0289 & 0.0355 & 0.0660 & 0.0994 & 0.0362 & 0.0448 & 0.0438 & 0.0669 & 0.0239 & 0.0297 \\
MICRO    & 0.0570 & 0.0905 & 0.0310 & 0.0406 & 0.0675 & 0.1026 & 0.0365 & 0.0463 & 0.0496 & 0.0743 & 0.0264 & 0.0332 \\
SLMRec   & 0.0540 & 0.0810 & 0.0296 & 0.0361 & 0.0676 & 0.1007 & 0.0374 & 0.0462 & 0.0452 & 0.0675 & 0.0247 & 0.0303 \\
BM3      & 0.0538 & 0.0857 & 0.0301 & 0.0378 & 0.0659 & 0.0979 & 0.0354 & 0.0437 & 0.0450 & 0.0669 & 0.0243 & 0.0295 \\
\hline
DiffMM   & 0.0604 & 0.0942 & 0.0319 & 0.0406 & 0.0696 & 0.1039 & 0.0377 & 0.0462 & 0.0567 & 0.0848 & 0.0302 & 0.0372 \\
MCDRec   & \underline{0.0644} & \underline{0.1013} & 0.0343 & 0.0438 & \underline{0.0737} & \underline{0.1100} & \underline{0.0392} & \underline{0.0488} & \underline{0.0582} & \underline{0.0864} & \underline{0.0316} & \underline{0.0387} \\

\hline
\textbf{JBM-Diff}
& \textbf{0.0670*} & \textbf{0.1065*} & \textbf{0.0368*} & \textbf{0.0469*}
& \textbf{0.0750*} & \textbf{0.1148*} & \textbf{0.0410*} & \textbf{0.0513*}
& \textbf{0.0629*} & \textbf{0.0937*} & \textbf{0.0342*} & \textbf{0.0420*} \\
\hline
Improv.
& 4.04\% & 5.13\% & 5.44\% & 6.59\%
& 1.76\% & 4.36\% & 4.59\% & 7.17\%
& 8.07\% & 8.45\% & 8.23\% & 8.53\% \\
\hline

\end{tabular}
}
\end{table*}

\subsubsection{Experiment Settings}

To achieve fair comparison, we set the training batch size to 512 and the test batch size to 1024 for all models. The embedding size is 64. For all graph-based methods, the number of layers in the user-item collaborative interaction graph is set to 2. We use the Xavier method to initialize model parameters and optimize all models with the Adam optimizer, with a learning rate set to 0.001. The total number of epochs is set to 1000, and an early stopping strategy with 10 epochs is implemented. For the hyperparameter selection of this model, grid search is employed. Specifically, the time steps for the conditional denoising model are adjusted within \{5, 10, 15, 20\}; the k-nearest neighbor selection for the item-item affinity graph in the modality is chosen from \{5, 10, 15, 20\}; and the update degree w of modality denoising is adjusted within \{0.1, 0.3, 0.5, 0.7, 0.9\}. Regarding the weight distribution of different losses, $\mu_{dm}$, $\mu_{mm}$, and $\mu_{cl}$ are chosen from \{0.0001,0.0005,0.001,0.005,0.1,0.5\}. The two temperature coefficients $\lambda$ and $\gamma$ that control preference consistency and modality conflict are chosen from \{0.1, 0.2, 0.5, 1.0, 2.0, 3.0\}, respectively.


\subsection{Modal Performance(RQ1)}

Table ~\ref{table2} presents the comparison of recommendation performance between the JBM-Diff model and other baselines on three datasets. Based on the evaluation results, we draw the following conclusions:
\begin{itemize}[leftmargin=*] 
    \item The superiority of JBM-Diff's performance. JBM-Diff exhibits the best performance across all metrics on the three datasets. In terms of improvement, Recall@20 has increased by 5.13\%, 4.36\%, and 8.45\% on the three datasets compared to the second-best model, respectively. Furthermore, its performance significantly surpasses traditional collaborative filtering and graph recommendation models, demonstrating the effectiveness of its multimodal recommendation.
    \item The effectiveness of behavior guidance in multimodal noise reduction. Compared with many multimodal graph recommendation models based on contrastive learning, JBM-Diff demonstrates significant performance improvement. This is because previous work overlooked the noisy nature of modal features themselves. In contrast, our work utilizes a conditional noise reduction model, using behavior features as conditions, to denoise multimodal features and remove information unrelated to user preferences, thereby further enhancing recommendation performance.
    \item The effectiveness of behavior debiasing based on modality preference consistency. Compared with many multimodal recommendation models based on diffusion models, JBM-Diff still performs well. This is because we not only consider enhancing behavior graph representation learning by utilizing multimodal features, but also focus on the impact of the partial order accuracy of training sample pairs on recommendation performance. Through consistency verification of modality preference and behavior preference, as well as conflict detection between modality preferences, we set a confidence level for each sample pair, alleviating the impact of behavior noise from the perspective of data augmentation, thereby enhancing recommendation performance.
\end{itemize}




\subsection{Ablation Study(RQ2)}
To comprehensively explore the influence of various factors, we conducted an ablation study on the core modules of JBM-Diff:

\begin{itemize}[leftmargin=*] 
    \item \textbf{w/o Multi-Modal Denoising (MMD)}: We removed the behavior-guided multi-modal denoising module and directly input the preprocessed multi-modal features into the corresponding item-item affinity graph.
    \item \textbf{w/o Feature Fusion (FF)}: We removed the fusion logic of behavioral and semantic features in the multi-view message propagation and feature fusion module, and instead directly used the average pooling results of behavioral features and features from each modality as the final embedding.
    \item \textbf{w/o Behavior Debiasing (BD)}: We removed the behavior debiasing module based on modality consistency, meaning we did not weight the training sample pairs, and used the original BPR loss as the primary training objective for training.
\end{itemize}



\begin{table}[htbp]
\centering
\caption{Ablation study on core components of JBM-Diff.}
\label{table3}
\begin{tabular}{llcccc}
\toprule
Datasets & Variants & R@10 & R@20 & N@10 & N@20 \\
\midrule

\multirow{4}{*}{Baby}
& JBM-Diff & 0.0670 & 0.1065 & 0.0368 & 0.0469 \\
& w/o MMD & 0.0282 & 0.0474 & 0.0152 & 0.0201 \\
& w/o FF  & 0.0461 & 0.0758 & 0.0242 & 0.0318 \\
& w/o BD  & 0.0653 & 0.1026 & 0.0355 & 0.0451 \\

\midrule

\multirow{4}{*}{Sports}
& JBM-Diff & 0.0750 & 0.1148 & 0.0410 & 0.0513 \\
& w/o MMD & 0.0325 & 0.0492 & 0.0178 & 0.0222 \\
& w/o FF  & 0.0449 & 0.0707 & 0.0245 & 0.0312 \\
& w/o BD  & 0.0719 & 0.1094 & 0.0392 & 0.0489 \\

\bottomrule
\end{tabular}
\end{table}

\begin{figure*}[t]
    \centering
    \includegraphics[width=1\linewidth]{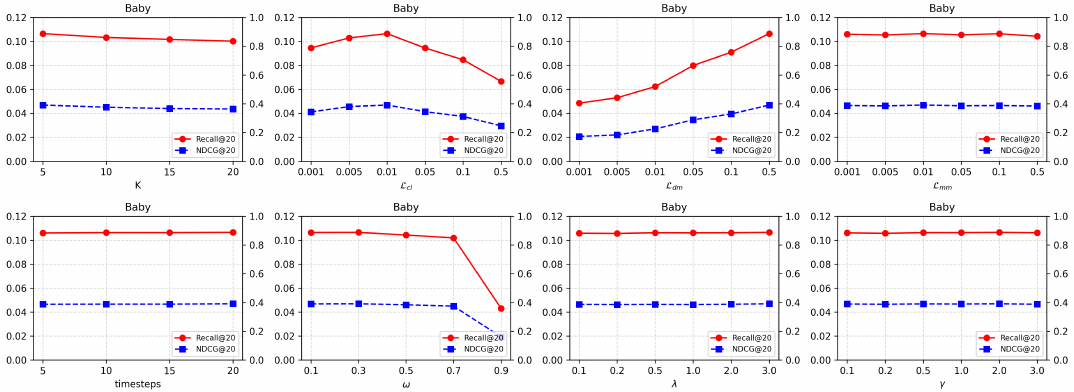}
    \caption{Performance comparison of different hyperparameters under various values.}
    \label{hyper}
\end{figure*}

The experimental results of JBM-Diff and its variants on the Baby and Sports datasets are presented in Table \ref{table3}. From these results, we observe the following conclusions: 1) When the multimodal denoising module is removed, the recommendation performance significantly decreases, even inferior to that relying solely on collaborative features for recommendation. This indicates the presence of severe noise in multimodal features and underscores the necessity of using behavior vectors to guide the removal of redundant information in modal features that is unrelated to user preferences. 2) After removing the feature fusion module, the recommendation performance also shows a noticeable downward trend, indicating the existence of a semantic gap between modal features and collaborative features. Simple average pooling cannot effectively align the spatial gap between the two, which indirectly confirms the importance of our alignment task. 3) When the behavior debiasing module is removed, the recommendation performance also decreases, indicating that there are cases in the training sample pairs that contradict the user's true preferences. By assigning weights to the partial order relationship of each sample pair, we effectively avoid behavior noise caused by systematic errors, thereby further improving recommendation accuracy.

\subsection{Hyperparameter Study(RQ3)}
In this section, we will explore the sensitivity of JBM-Diff to some key hyperparameters on the baby dataset. The relevant experimental results are shown in Fig. ~\ref{hyper}.

\subsubsection{Effects of the number of item neighbor $K$.}
To avoid propagating information from unrelated items, we constructed an item-item affinity graph using only the K most similar items. Experiments show that K=5 yields the best results, and as K increases, the recommendation performance of JBM-Diff continuously decreases. This indicates that aggregating too many semantic neighbors introduces unnecessary noise. Therefore, in practical applications, it is not advisable to set K to a too large value.

\subsubsection{Effects of the align loss weight $\mathcal{L}_{cl}$.}
We designed a self-supervised contrastive learning task during multi-view feature fusion to align behavioral and semantic features. Here, we explored the value of $\mathcal{L}_{cl}$. Experimental results showed that the best performance was achieved when $\mathcal{L}_{cl}=0.01$. If the weight is too small, the learning effect of the alignment task is poor, affecting the final feature representation; if the weight is too large, it may affect the training of other objectives.

\subsubsection{Effects of the diffusion loss weight $\mathcal{L}_{dm}$.}
We conducted an exploration on the value of the diffusion loss weight $\mathcal{L}_{dm}$ in the conditional diffusion model for multimodal noise reduction. Experimental results indicate that when $\mathcal{L}_{dm}$ is too small, the training effect of the diffusion model is unsatisfactory, leading to poor recommendation performance. This further underscores the necessity of multimodal noise reduction.

\subsubsection{Effects of the inter-modality align loss weight $\mathcal{L}_{mm}$.}
We conducted an exploration on the value of the contrastive loss weight $\mathcal{L}_{mm}$ for the inter-modal alignment task in multimodal noise reduction. Experimental results indicate that $\mathcal{L}_{mm}$ does not significantly affect the performance of JBM-Diff. This may be attributed to the relatively balanced spatial distribution of modal features in the selected dataset, where different modalities exhibit a relatively consistent understanding of user preferences.

\subsubsection{Effects of the steps of diffusion model $timesteps$.}
We conducted an exploration on the recommended time step size for the diffusion model. Experimental results indicate that the size of the time step has a negligible impact on model performance, and a small number of steps can achieve good denoising effects. This is because our denoising model has a small number of trainable parameters, indicating that our model can achieve high performance with relatively low complexity.

\subsubsection{Effects of the update levels of multi modal features $\omega$.}
After the noise reduction processing through the conditional diffusion model, we control the degree of updating of modal features through $\omega$. The smaller $\omega$ is, the more original features are retained. Experimental results show that as $\omega$ increases, the model performance first increases and then decreases, and if $\omega$ is too large, the model performance will significantly decline. This illustrates the importance of retaining original features.

\subsubsection{Effects of the temperature of behavior debiasing $\lambda$ and $\gamma$.}
We analyze the impact of the temperature hyperparameters $\lambda$ and $\gamma$, which control the weights of preference consistency and modality conflict, respectively. Experimental results show that the model performance remains relatively stable across different combinations of $\lambda$ and $\gamma$. This indicates that the proposed debiasing mechanism is robust to the choice of these hyperparameters. We attribute this stability to the complementary roles of preference consistency and modality conflict, which naturally balance each other and enable adaptive sample weighting without requiring delicate tuning.


\subsection{Performance on Various Noisy Scenarios}

To demonstrate the robust anti-noise capability of our model, we designed corresponding noise injection experiments on multimodal semantics and behavior interaction. Furthermore, we compared our model with the following baseline models: 1) Traditional graph recommendation model: LightGCN~\cite{he2020lightgcn} 2) Multimodal recommendation model based on hypergraph: LGMRec~\cite{guo2024lgmrec} 3) Multimodal recommendation model based on contrastive learning: LATTICE~\cite{zhang2021mining}, MMGCN~\cite{wei2019mmgcn} 4) Multimodal recommendation model based on diffusion model: MCDRec~\cite{ma2024multimodal}.

\begin{figure}[t]
    \centering
    \includegraphics[width=1\linewidth]{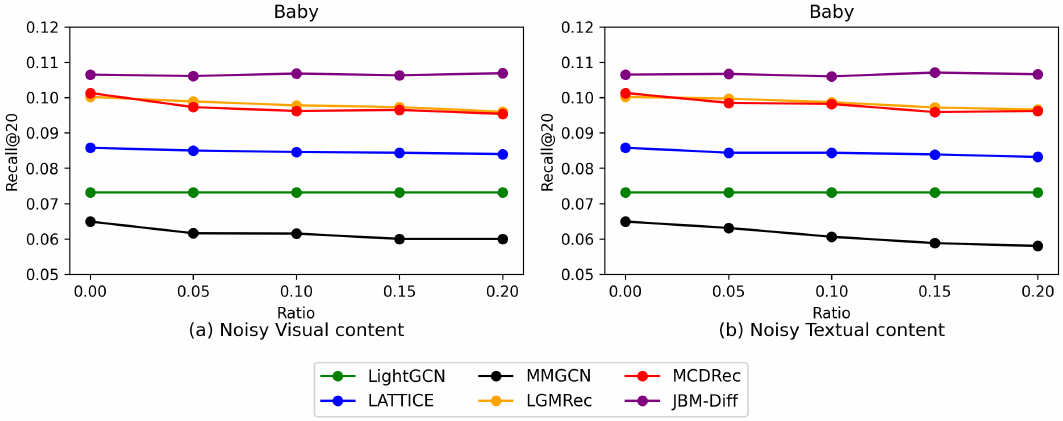}
    \caption{Performance in various noisy multi-modal content scenarios on Baby dataset}
    \label{noisy_modal}
\end{figure}

\begin{figure}[t]
    \centering
    \includegraphics[width=1\linewidth]{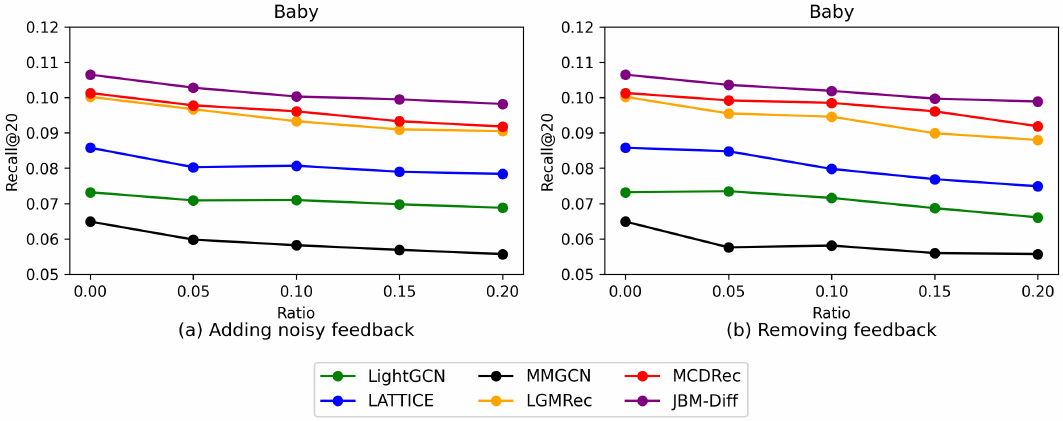}
    \caption{Performance in various noisy user feedback scenarios on Baby dataset}
    \label{noisy_feed}
\end{figure}

\subsubsection{Performance on Noisy multi-modal content.}
To demonstrate the denoising capability of JBM-Diff for multimodal content, we intentionally introduced noise into the multimodal content of each dataset and evaluated the recommendation performance. To construct a dataset with more noise in multimodal content, we first randomly sampled from the training set. For each sampled item i, we randomly selected another item j $(j \neq i)$ from the training set and replaced the original modal features of item i with the features of item j. We randomly replaced 5\%, 10\%, 15\%, and 20\% of the item modal features in the dataset. The replacement ratio was limited to 20\% to prevent excessive noise from causing anomalies in all modalities. We replaced either visual or textual patterns. Each time, we only modified the features of one modality while keeping the feedback data unchanged.


The experimental results are shown in Fig. ~\ref{noisy_modal}, from which we can observe that regardless of the modality to which noise is added, the JBM-Diff model outperforms other baseline models in recommendation performance under different noise ratios. Furthermore, as noise increases, the model performance remains unaffected, further validating the effectiveness of the behavior-guided multimodal denoising module.

\subsubsection{Performance on Noisy feedback.}
To demonstrate the ability of JBM-Diff to denoise user feedback, we contaminated the feedback data in two ways. (1) Randomly generate interactions that do not exist in the dataset and add them to the training set. (2) Randomly remove interactions from the training set. We only modified the training set, keeping the validation and test sets unchanged. Similarly, to generate noise scenarios and avoid excessive noise, we adjusted 5\%, 10\%, 15\%, and 20\% of the user feedback in the training set.


The experimental results are shown in Fig. ~\ref{noisy_feed}. It can be observed that, whether noise is added to the interaction data or some interaction data is removed, the recommendation performance of the JBM-Diff model remains superior to other baseline models, and it is relatively less affected by noise and data sparsity. This is because we utilize modal features to set weights for the partial order consistency of each training sample pair, assigning lower confidence to potentially noisy samples, thereby reducing the impact of behavioral noise on performance. Compared to traditional graph recommendation models like LightGCN, our model mitigates the impact of data sparsity on performance by incorporating multimodal features as side information.

\section{Conclusion}

Addressing the issues of modal noise and feedback bias in multi-modal recommendation, we propose a conditional graph diffusion model, JBM-Diff, based on behavior guidance and modal consistency, to jointly denoise multi-modal features and user feedback. The innovation of this work lies mainly in two aspects. Firstly, we introduce a behavior-guided conditional diffusion model for each modal to remove redundant information unrelated to user preferences. Secondly, we utilize multi-modal features to examine preference consistency and set credibility for sample pairs to achieve data augmentation. Furthermore, various experiments designed on three open-source datasets verify the effectiveness of our model.
\bibliographystyle{ACM-Reference-Format}
\bibliography{references}

\appendix









\end{document}